\documentclass[%
 reprint,
 amsmath,amssymb,
 aps,
prd,
floatfix,
]{revtex4-2}

\usepackage{graphicx}
\usepackage{dcolumn}
\usepackage{bm}

\begin{document}

\title{BUFF: Boosted Decision Tree based Ultra-Fast Flow matching}

\author{Cheng Jiang}
\email{C.Jiang-19@sms.ed.ac.uk}
 \affiliation{%
 School of Physics and Astronomy, University of Edinburgh, EH9 3FD, Edinburgh, United Kingdom
}%
\author{Sitian Qian}%
 \email{stqian@pku.edu.cn}
\affiliation{%
 School of Physics, Peking University, 100871, Beijing, China
}%

\author{Huilin Qu}
\email{huilin.qu@cern.ch}
\affiliation{
 CERN, EP Department, CH-1121 Geneva 23,  Switzerland
}

\date{\today}

\begin{abstract}
    Tabular data stands out as one of the most frequently encountered types in high energy physics. Unlike commonly homogeneous data such as pixelated images, simulating high-dimensional tabular data and accurately capturing their correlations are often quite challenging, even with the most advanced architectures. Based on the findings that tree-based models surpass the performance of deep learning models for tasks specific to tabular data, we adopt the very recent generative modeling class named conditional flow matching and employ different techniques to integrate the usage of Gradient Boosted Trees. The performances are evaluated for various tasks on different analysis level with several public datasets. We demonstrate the training and inference time of most high-level simulation tasks can achieve speedup by orders of magnitude. The application can be extended to low-level feature simulation and conditioned generations with competitive performance.

\end{abstract}

\maketitle

\section{Introduction}\label{sec1}

The collisions happening at the Large Hadron Collider (LHC) \cite{lhc}, following the nature of quantum mechanics, are intrinsically probabilistic. This necessitates a vast number of events to acquire meaningful information that can validate theoretical predictions. Oftentimes, we rely on the large-scale Monte Carlo (MC) event production \cite{mc} with a dedicated, and probably lengthy production pipeline to perform complicated calculations of the Standard Model. In the upcoming High-Luminosity LHC (HL-LHC) era \cite{hilhc}, it is expected that the trend for these simulations will shift towards a more CPU-intensive environment \cite{mcchallenge}, given the much larger amount of data-taking. It is crucial to prevent the possible limitation in physics output due to insufficient statistics arising from constraint computational budget.

The successful application of deep generative models as fast surrogate models in collider physics allows for the rapid generation of particle collision data, drastically reducing event generation period without sacrificing accuracy. Diffusion models (DM) \cite{score,ddpm,diffusionbeatgan} excels in generating high-quality, detailed samples by iteratively denoising random noise into clean output, showcasing impressive capabilities in tasks like fast calorimeter simulation \cite{calodiffu,caloscore,caloscore2,calocloud2,directdiff,pcjedi}. One common challenge of those DM applications is that they only reach the ideal performance after relatively long generation time. Many studies adopt the novel distillation techniques with pre-trained model such as progressive distillation and consistency model \cite{pd,cm,icm}. On the other hand, normalizing flows (NFs) \cite{nfreview,inn} stand out for their ability to model complex distributions with a small amount of time, making them especially powerful in fields requiring precise probability density estimation. Different generative models have emerged in various applications \cite{calogan,calom,caloqvae,caloshowergan,deeptreegan}. For the MC event generator, several applications focus on conditional phase space sampling \cite{madnis,madnis2,zunis,mm,nfeventgen,iflow,cqmnf}. Flow-based model could also produce good quality of shower rapidly \cite{caloflow,caloflow2,calopointflow2,l2flow,supercalo}. Those models are applied extensively to handle tasks such as unfolding \cite{cinnfold,sbfold}, shower shape correction, anomaly detection \cite{ad1,adben,nfad1,treead1} and quark gluon tagging \cite{qg}. 

Whereas, NF intrinsically can not reach desirable performance once moving to higher dimensions \cite{glow,fm}. Furthermore, most architectures adopted into high energy physics (HEP) are tailored for image synthesis. The majority of quantities or variables needed at the analysis level of event content are designed to be sensitive to certain phase spaces in one channel or aim to explore the deeper structure of hadronic objects. These variables have shapes far more complex than a simple Gaussian and are often treated as tabular data. Current approaches, which utilize the most advanced architectures to simulate tabular data in high dimensions and precisely capture their correlations, can be notably challenging, especially when considering the significantly longer training and validation times required for each update in analysis.

Recently, a new type of generative modelling class was introduced, inspired from the score matching diffusion model \cite{score} and optimal transport \cite{ot}, conditional flow matching (CFM) \cite{cfm} offers a simulation-free approach to directly match the vector field, demonstrating good scalability to very high dimensions. In this work, we used several techniques mentioned from the study \cite{flowbdt} to replace the traditional Neural Network backbone CFM with Gradient Boosted Trees (GBT) \cite{bdt,gbt}. We named the framework Boosted Decision Tree based Ultra-Fast Flow matching (BUFF) with the model flowBDT. The reasonale behind using tree-based CFM is that the vector fields and conditional probability paths generated from N-dimensional tabular data should also hold the similar structures inherent to them. Several comprehensive studies \cite{bdtbetter1,bdtbetter2} indicate that tree-based models surpass most of deep learning models in performance for tasks specific to tabular data, such as classification and regression. In Section~\ref{sec2}, we delve into the workings of the Tree-based Conditional Flow Matching and share some improved understanding of this method, alongside its novel applications. Section~\ref{sec3} describes the datasets utilized in our study. In Section~\ref{sec4}, we present the performance outcomes across various tasks using diverse datasets and a range of comprehensive metrics. Lastly, Section~\ref{sec5} offers potential applications and future directions for this line of research.

\section{Improved Understanding for Tree-based Flow matching}\label{sec2}

Suppose $f: \mathbb{R}^d \rightarrow \mathbb{R}^d$ is a continuous differentiable map between target distribution $p_1$ and base distribution $x \sim q_0$, which usually is multidimensional Gaussian. Normalizing flow (NF) requires each composition of the inverse map $f(x) = f_{T-1}\circ...\circ f_{0}(x)$ to have valid Jacobian determinants during sampling phase. By the change-of-variable rules, the log-likelihood loss for NF training is written as \cite{glow}:

\begin{subequations}
\begin{equation}
\begin{aligned}
    p_1(x_T) = q_0(f^{-1}(x_T)) \hspace{1mm} |\det[\frac{\partial f^{-1}}{\partial x_{t}}(x_T)]|\\
    = \frac{q_0(x)}{|\det[\frac{\partial f}{\partial x}(x)]|}, \hspace{4mm} x = f^{-1}(x_T)
\end{aligned}
\end{equation}
\begin{eqnarray}
    \log p_1(x_T) = \log q_0(x) - \sum_{t=0}^{T} \log |\det[\frac{\partial f_t}{\partial x_{t}}]|
\end{eqnarray}    
\end{subequations}

From the formulation for the training loss, there are two drawbacks for transitional NF: the strict requirement for invertible map $f$ and inefficient Jacobican determinant calculation. These would make NF predicts worse when scaling up to very high dimensions. Later, continuous normalizing flow (CNF) is proposed to replace the second term in Equation.2 using the property that the trace of the Jacobian matrix is equivalent to the log probability of the continuous random variable: 

\begin{eqnarray}
    \log p_1(x_T) = \log q_0(x) - \int_{t=0}^{T} \mathrm{Tr}(\frac{\partial f}{\partial x_{t}}) \mathrm{d}t
\end{eqnarray}

This allows CNF to have a more efficient training stage given trace is generally easier to solve. However, the vector field from a flow is still integrated over time ( i.e. $\frac{\partial x}{\partial t} = f(x_t,t)$ ). The computational demanding training and bad scalability problems insist in these  simulation-based approaches.

Flow matching proposes a simulation-free loss function which directly matches the probability path $p_t(x)$ and vector field $u_t(x)$ for time steps $t$ uniformly distributed between 0 and 1. For any learnable vector field $v_t(x)$ parameter $\theta$ from a model with stochastic regression objective, the loss defined as \cite{fm}:
\begin{eqnarray}
    \mathcal{L}_{FM}(\theta) = \mathbb{E}_{t,p_t(x)}||v_{\theta}(t,x)-u_t(x)||^2
\end{eqnarray}
To compute this loss function needs the explicit knowledge of the target vector field $u_t$. One could use the conditioning variable $z$ to define the CFM objective \cite{cfm}: 
\begin{subequations}
\begin{eqnarray}
    p_t(x) = \int p_t(x|z)q(z)\mathrm{d}z
\end{eqnarray}
\begin{eqnarray}
    \mathcal{L}_{FM}(\theta) = \mathbb{E}_{t,q(z),p_t(x|z)}||v_{\theta}(t,x)-u_t(x|z)||^2
\end{eqnarray}
\end{subequations}
Both loss functions should hold the same gradient w.r.t. weights $\theta$.

There exists a specific, non-universal scenario in which GBT can be utilized here. Since GBT classes are non-differentiable, they cannot be directly applied in generative models such as Generative adversarial networks \cite{gan0} or variational autoencoders \cite{vae0}, which rely on composing multiple models during training. In typical neural network-based flow matching, we employ Stochastic Gradient Descent (SGD) \cite{sgd} with random sampling to minimize the expectations from Equation.4 and Equation.6 over a mini-batch. Additionally, we adapt the setup for GBT to 'mimic' batch training as practiced in deep learning models. The data points are duplicated repeatedly to approximate the expectation calculated over all possible pairs between input and target sample. We use the linear interpolation and independent conditional flow matching to build the training dataset of vector fields at each noise level, there will be different interpolant class for specific objectives \cite{cfm,sbcfm}.  A small individual GBT model with 3-4 depth and 50-100 estimators is used on each time step. The training time is directly proportional to both the tree depth and number of estimators in one GBT model. Compared with other deep learning model, the extremely fast training and inference time of those light GBT model can provide a robust performance with only fraction of time for high-level physics task. 

Since different models are trained for each noise level, the number of time steps during inference is fixed. Nevertheless, we retain the flexibility to adjust the sampler for solving ODE. The crucial aspect that needs investigation is whether those different models can predict with equivalent accuracy for ODE calculations with high-order numerical methods across various time steps. In our research, we modify the original setup: instead of employing the first-order Euler solver, we evaluate the performance using the second-order Midpoint solver, as well as higher-order solvers such as Runge-Kutta \cite{rk}, Tsitouras, and Dormand-Prince \cite{dpori}. The solvers that best fit our tasks are identified as the Midpoint solver at 30-40 steps and the Dormand-Prince solver at 15-30 steps. Consequently, we have selected these two solvers as our defaults. In addition to this, we find that varying the number of output on each tree of the model significantly reduces the sampling time. We keep those hyperparameters small enough to achieve decent performance with just negligible time typically required.

\section{Dataset}\label{sec3}

\subsection{JetNet}\label{subsec3.1}
Around 200,000 events of \textit{JetNet} dataset \cite{jetnet} are simulated at leading order using \verb+Madgraph+ \cite{madgraph}. Each of them focus on a relatively narrow kinematic range so that relative transverse momenta of the generated parton and gauge boson are within a small window. Those parton-level events undergo the hadronization process via \verb+Pythia+ \cite{pythia}. The transverse momentum ($p^\mathrm{T}$) of particle jets is around 1 TeV. Jets are clustered by anti$-k_\mathrm{T}$ algorithm with a cone size of $R =$ 0.4. 

The first 30 particles with highest $p^\mathrm{T}$ inside the jets are taken into account. Four particle-level variables are available: the relative location w.r.t. jets on angular coordinate frame: $\eta^\mathrm{rel} = \eta^\mathrm{particle} - \eta^\mathrm{jet}$, $\mod{(\phi^\mathrm{rel} = \phi^\mathrm{particle} - \phi^\mathrm{jet}, 2 \pi)}$, relative transverse momentum of the particle defined by $p^\mathrm{rel}_\mathrm{T} = p^\mathrm{particle}_\mathrm{T}/p^\mathrm{jet}_\mathrm{T}$. To distinguish the jets with fewer than 30 particles, fourth variable is tailored to tell if the particle is genuine or zero-padded. 

Different types of jets are provided in the dataset, either from gluons, light or top quarks, or vector bosons. The gluon and light quarks usually have simple topology. We choose top quark jets to derive the high level features as they should have the complex topology to capture the inner correlations among subjets and structural distributions for the global point cloud. 

\subsection{CaloChallenge}\label{subsec3.2}

The Fast Calorimeter Simulation Challenge 2022 \cite{calochallenge} aims to create and evaluate a rapid and highly accurate calorimeter simulation by benchmarking deep learning methods. The datasets use the GEANT4 \cite{GEANT4:2002zbu} framework to simulate different kinds of particle showers inside either a irregular or concentric calorimeter.

In each dataset, showers are simulated to travel along the z-axis within a cylindrical detector. Cells in the radial $r$ and angular $\alpha$ bins are arranged on each sampling layer of the calorimeter. \textit{CaloChallenge} dataset 1 contains photon shower simulated by \verb+Geant4+ with 368 voxels on 5 different calorimeter layers. Layer 1 and 2 are designed to have more granular layout than other three layers. By assigning the different voxel energies and locations according to the specific $r$ and $\alpha$ bins, one can reconstruct the total deposited energy of the shower for the particle with incident energies from 256 MeV up to 4 TeV. Even though the other two \textit{CaloChallenge} datasets are more granular than the first dataset, dataset 1 provides a better shower description within irregular geometry, and also a smoother translation to tabular data.

\subsection{Jet Datasets for Unfolding}\label{subsec3.3}

The dataset used for unfolding task as described in Section.~\ref{subsec4.3} is from \cite{omnifold}. It contains QCD jets from Z + jets events that are simulated by either \verb+Herwig+ \cite{herwig} or \verb+Pythia+ with different tunes and parton distribution functions. The jets are clustered by anti$-k_T$ algorithm within $R =$ 0.4. We consider event to have the leading jet with a Z boson $p^\mathrm{T} >$ 200 GeV to make the event topology unaffected by the limited acceptance. Multiple datasets is required to perform the unfolding between full particle phase spaces and particle flow objects. The detector distortions are simulated by fast detector simulator of the CMS detector through \verb+Delphes+ \cite{delphes} with concrete particle flow reconstruction. Six different quantities derived from the substructure of the leading jet are used to evaluate the performance.

\subsection{Schr{\"o}dinger Bridge Refinement}\label{subsec3.4}

Schr{\"o}dinger Bridge Refinement involves two distinct datasets of electron particles traveling through an electromagnetic calorimeter. The conditional input dataset \verb+GFlash+ \cite{sbrefine,gflash1,gflash2} approximates electron showers using a parameterized fast simulation. The formulation is based on the knowledge of longitudinal and radial showers in the sampling calorimeter. We keep the Moli{\'e}re radii and radiation length as their default values. Additionally, the \verb+GFlash+ procedures account for the effects of sampling geometry and materials on shower shapes.

Refining showers from the fast simulation to match the full simulation target from \verb+Geant4+ requires an injective mapping between them. Therefore, we ensure that the incident particle energies of both simulations remain the same for each training event. The electron shower has a regular size of $10 \times 10$ calorimeter cells with fixed incident energies of 50 GeV. The core idea for this task is to use the approximate shower as input conditions which should contain more comprehensive information about geometry and material.

\begin{figure*}[htbp]
    \centering
    \includegraphics[width=\textwidth]{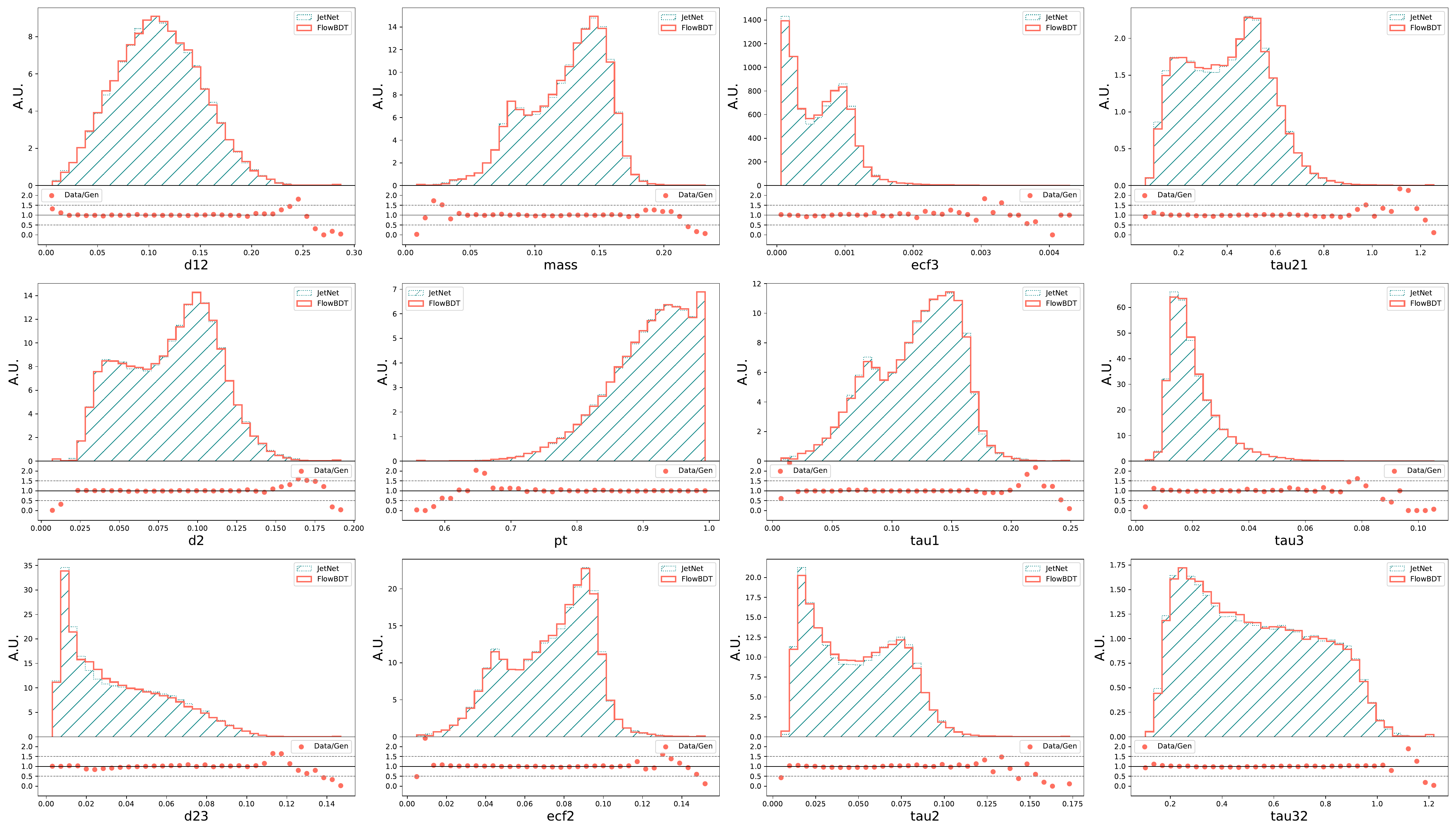}
    \caption{The histogram and ratio plots for generated simultanously and original jet variables. Shaded: from JetNet, solid line: from flowBDT}
    \label{fig1}
\end{figure*}

\section{Tasks}\label{sec4}

Based on the new design and knowledge improvement of the model, various potential HEP applications will be shown in this section. We first demonstrate the baseline performance task for end to end fast simulation as done with traditional normalizing flow \cite{flashsim} in section~\ref{subsec4.1}. Then, section~\ref{subsec4.2} shows the advantages of flow matching and GBT model by moving to a much higher dimension regime. flowBDT can still give decent performance for simulating hundreds of cells within irregular geometry and particles inside jets. Section~\ref{subsec4.3} illustrates the model is also fully capable to do conditioned generation and distribution mapping from non-tractable prior distribution. 

\subsection{End to end fast simulation}\label{subsec4.1}

Most of the high level variables used in analysis-level are derived via a long simulation chain. Those variables are chosen among a larger subset of variables after digitization and reconstruction. Having sufficient statistics for those high-level variables is crucial for many analyses, while deriving those quantities starting from the generator level would be computationally demanding. The possible approach is to directly generate those high-level observables according to the information from full simulation. Skipping those intermediate steps would reduce the simulation time significantly. 

Whereas many variables used in the analysis have complicated distribution, the generative model has to learn to simulate a N dimensional distribution far away from pure Gaussian form. Capturing the intricate shapes and correlations of these distributions poses a considerable challenge. We studied the specific case by simulating four vector, substructure, and correlation function variables of highly boosted objects such as top jets. The total number of 12 variables calculated from 3 subjets of pseudo-jets by \textit{JetNet} are considered, specifically, The transverse momentum and mass of the jet $p^\mathrm{T}$, $m$, the N-subjetiness for exclusive n-jet $\tau_1$, $\tau_2$, $\tau_3$, the N-subjetiness ratio $\tau_{21} = \frac{\tau_2}{\tau_1}$, $\tau_{32} = \frac{\tau_3}{\tau_2}$, Energy splitting function defined by the splitting scale for n-jet exclusive clustering $d_{12}$, $d_{23}$, n\textsuperscript{th} order Energy correlation function $ecf_2$, $ecf_3$ calculated from $p^\mathrm{T}$ and $\Delta R$, finally $d_2 = \frac{ecf_3 \sum p^\mathrm{T}}{ecf_2^2}$.

The Dormand-Prince and Midpoint solver within 20-30 steps could already give satisfactory results for generating 100k events with all variables. Fig.\ref{fig1} suggests most of variables with complicated shapes have great agreement between target and generated sample. The best performance on histogram is seen for substructures and functions related to first two subjets, especially distributions with two asymmetric peaks. Despite a slight misalignment in the histograms of the last subjet, flowBDT can efficiently handle well for the tasks around 10-20 dimensions with both few-steps low and high order ODE solver.  
\begin{figure*}[htbp]
    \centering
    \includegraphics[width=\textwidth]{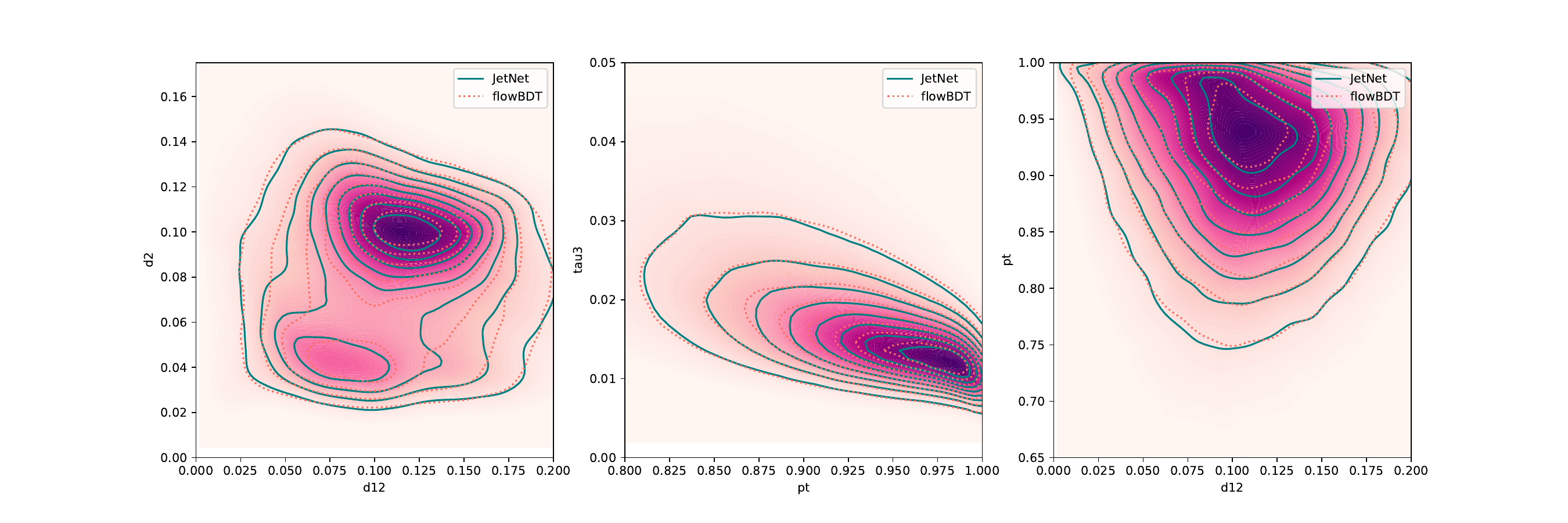}
    \caption{The kernel density estimation contour plot for generated and original 100k jet variables. Solid: from JetNet, dotted: from flowBDT}
    \label{fig2}
\end{figure*}

\begin{table}[htbp]

\begin{ruledtabular}
        \begin{tabular}{ccc}
        \textbf{Metrics} & \textbf{Sep power}$\times$100 & \textbf{EMD}$\times$10  \\
        \hline
        \textbf{$d_{12}$} & 0.059 & 0.0030  \\
        \textbf{$d_2$} & 0.14 & 0.0037 \\
        \textbf{$d_{23}$} & 0.0057 & 0.095 \\
        \textbf{$\tau_{1}$} & 0.096 & 0.042\\
        \textbf{$\tau_{2}$} & 0.15 & 0.0070\\
        \textbf{$\tau_{3}$} & 0.065 & 0.009\\
        \textbf{$mass$} & 0.086 & 0.0030 \\
        \textbf{$p^\mathrm{T}$} & 0.051 & 0.0085 \\ \textbf{$\tau_{21}$} & 0.045 & 0.020\\
        \textbf{$ecf_2$} & 0.084 & 0.0027 \\ 
        \textbf{$ecf_3$} & 0.0001 & 0.15\\
        \textbf{$\tau_{32}$} & 0.0097 & 0.031\\

       \end{tabular}

     \caption{Evaluation metrics for samples generated from flowBDT. The separation power is calculated from the difference of binned histograms between generated and original sample. Earth Mover's Distance, or Wasserstein Distance is derived from unbinned distributions.}
     \label{tab1}
\end{ruledtabular}
\end{table}

The evaluation metrics for binned and unbinned distribution for those variables are also provided in Table.~\ref{tab1}. Almost all the features sampled from the model give the impressive performance on f-divergence metric (i.e. triangular discriminator \cite{sep,tmva}). The Earth Mover's Distance \cite{emd0,wgan} maintains very low value for majority of generated features on unbinned distribution listed on Table.~\ref{tab1}. To do a comprehensive evaluation, variable simulated from our model should also have good pairwise correlation. We compare the correlation of generated sample with the original ones. Correlation contour plots are estimated by the Gaussian kernel. Representative correlations among four vector, substructure and energy splitting functions ($d_1$, $d_{12}$, $p^\mathrm{T}$, $\tau_3$) are shown in Fig.\ref{fig2}. Surprisingly, we find the model can have almost the same shape on every level of the contour lines, even for plots with multiple peak rings.

As demonstrated above, flowBDT generated sample in very good quality within negligible amount of inference time, 0.4-0.8 $ms$ for each event. Not only that, the training is incredibly fast for low dimensional tasks. We separate each time step with tiny BDT trees, trained on single thread of each CPU core. For 120k events, the training time is around 3-4 minutes per core. 

\subsection{Low level simulation for cells and jet constituents}\label{subsec4.2}

To fully explore the potential of our model, we increase the dimension of the input features to the next magnitude. In the study, we will focus on the dataset 1 which contains layers with different granularities. There are ways to map the features into a latent space, or treat them as flatten array to deal with irregular geometry. The latter does not usually take the shower as 3-D nor structured data. Following by this logic, all the voxel energies now are treated as tabular data. 

Output of GBT regression model by default need to be $(N,1)$, thus number of voxels per shower corresponds to the number of GBTs required for training. Instead, we choose to let each tree has one output so that the number of trees are reduced to only one GBT model per time step. This also makes 5-8 times speedup while inferencing with indistinguishable performance as default setting.
\begin{figure*}[htbp]
    \centering
    \includegraphics[width=\textwidth]{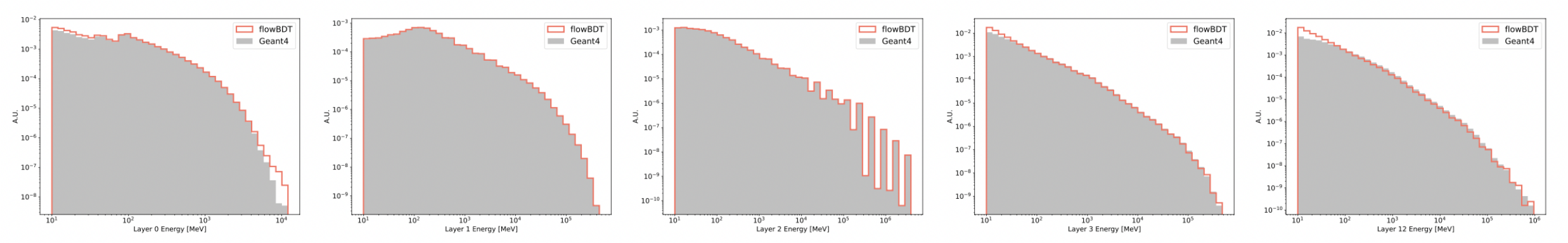}
    \caption{Different calorimeter layer energies of the photon sample generated by flowBDT and Geant4.}
    \label{fig3}
\end{figure*}

In Fig.~\ref{fig3}, we sum up the output raw voxels according to $\alpha$ and $r$ bins of different layers. The flowBDT predicts very well on more granular layers as layer 1 and 2. However, a shortfall in simulating this metric becomes apparent in the energy distribution tails of layer 0 and in instances where showers with low energy deposited on the last layer.

\begin{figure}[htbp]
    \centering
        \centering
        \includegraphics[width=0.6\linewidth]{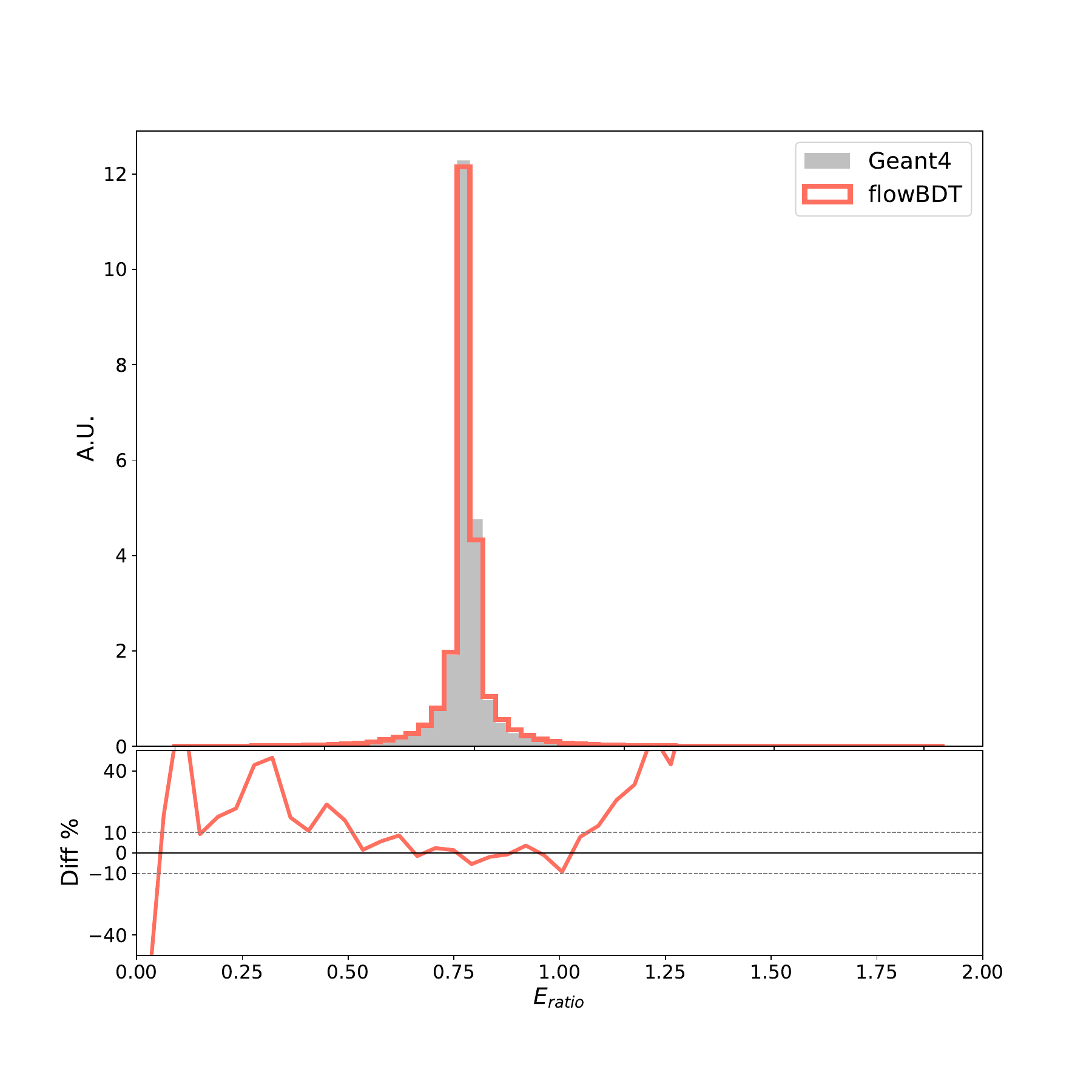}
        \caption{Shower response histogram and ratio plot of the photon sample generated by flowBDT and Geant4.}\label{fig4}
    
\end{figure}

\begin{figure}[htbp]
    \centering
        \centering
        \includegraphics[width=0.45\linewidth]{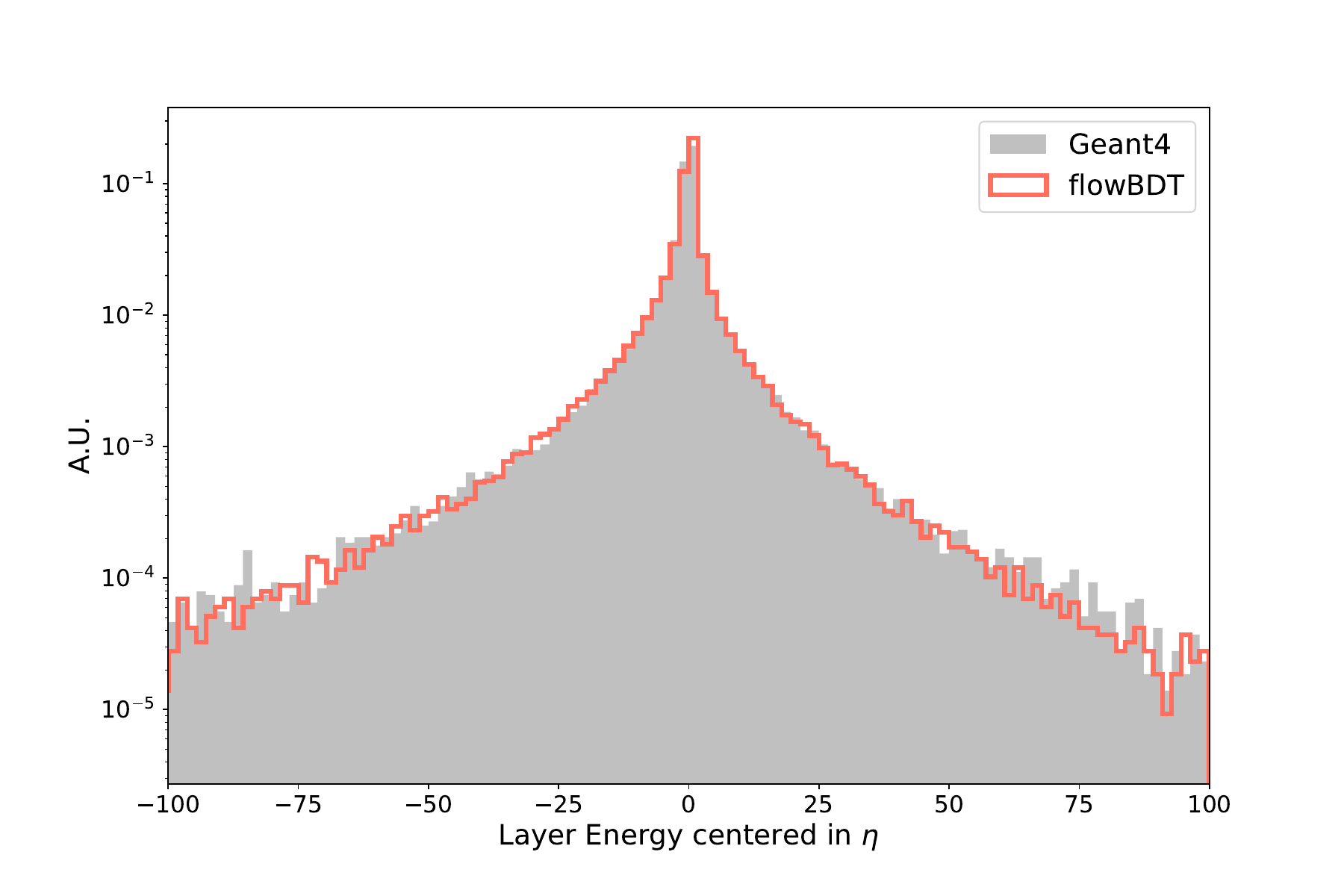} \quad
        \includegraphics[width=0.45\linewidth]{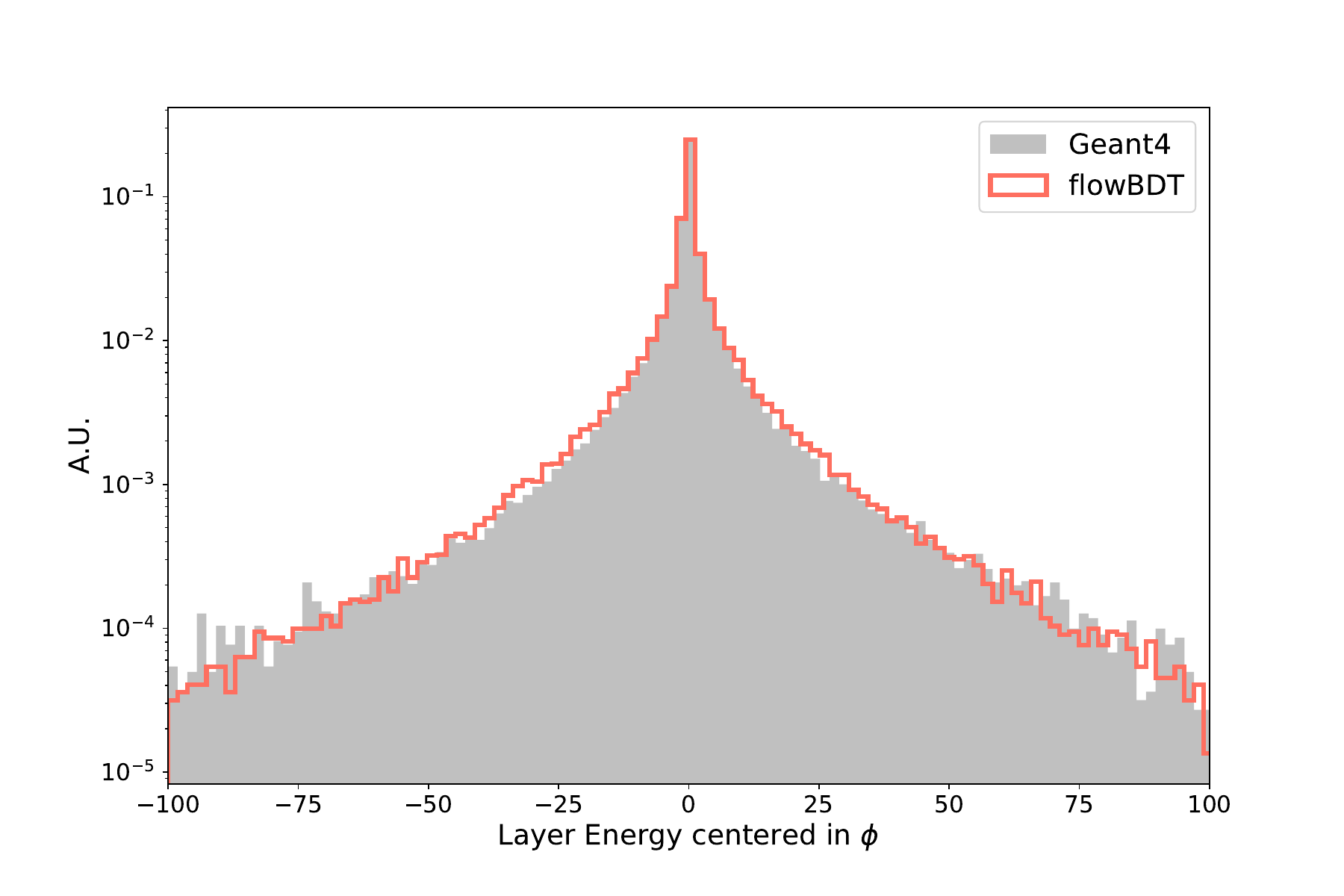}
        \caption{Center of energy in $\eta$ and $\phi$ direction on second layer of the photon shower generated by flowBDT and Geant4}\label{fig5}
    
\end{figure}

Shower response $E_{ratio}$ as an important quantity used in calibrating objects is shown in Fig.~\ref{fig4}. The $E_{ratio}$ generated by the model exhibits a shape and peak width closely mirroring those produced by Geant4. A similar correspondence is observed for another variable concerning the shower location, the center of energy in the $\eta$ and $\phi$ directions: $\Bar{\eta} = \frac{\langle \eta_i E_i \rangle}{\sum E_i}$ for cell location $\eta_i$, as illustrated in Figure.~\ref{fig5}.

\begin{figure*}[htbp]
    \centering

    \includegraphics[width=\textwidth]{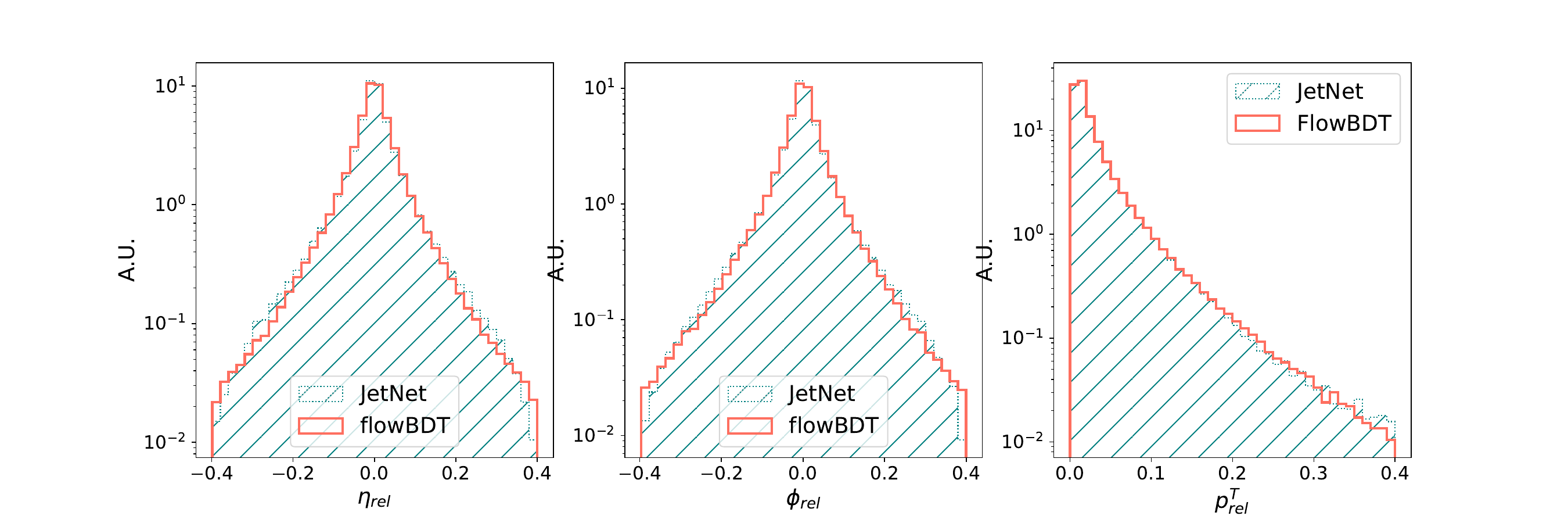}
    \caption{Relative location and $p^\mathrm{T}$ of the jet constituents inside gluon-initiated jets from flowBDT and JetNet.} 
    \label{fig6}
\end{figure*}

\begin{figure*}[htbp]
    \centering

    \includegraphics[width=\textwidth]{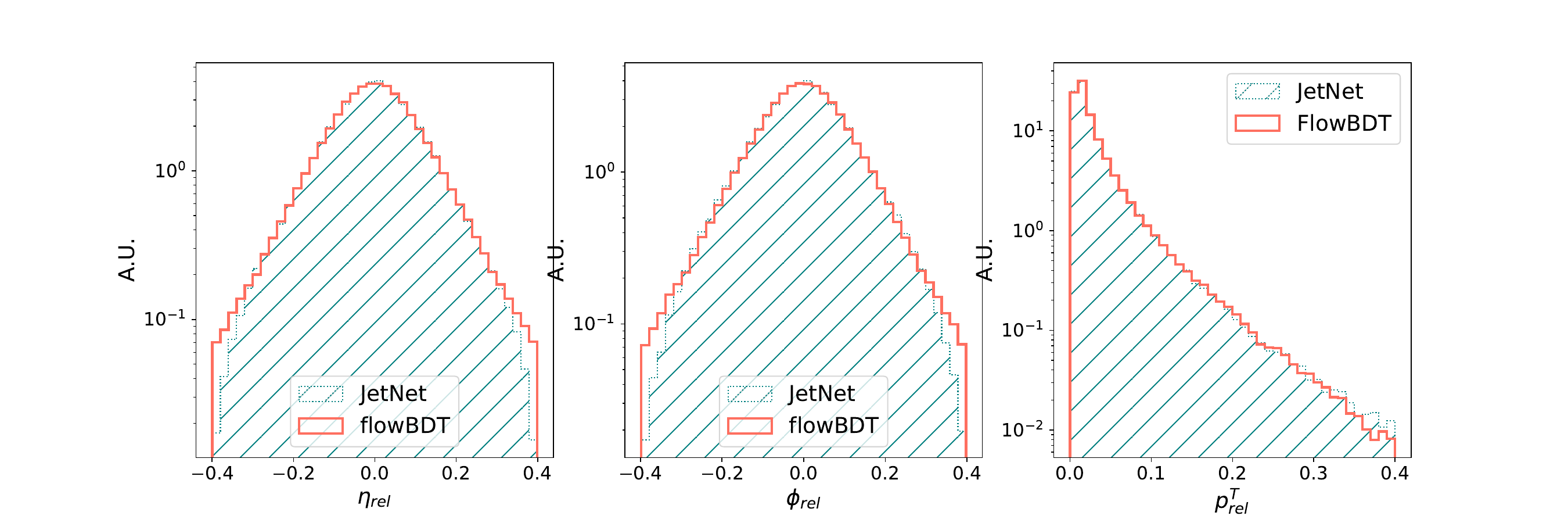}
    \caption{Relative location and $p^\mathrm{T}$ of the jet constituents inside top quark jets from flowBDT and JetNet.} 
    \label{fig7}
\end{figure*}

The same treatment is applied to the JetNet dataset. In this context, the goal is to regress the raw particle-level four-vector data from gluon-initiated and top jets. Generated relative $p^\mathrm{T}$ distribution has outstanding alignment with the histogram. A small deficiency was observed on the tails of relative location distribution compared to the original dataset, as illustrated in Figure~\ref{fig7}. This discrepancy means the limitations of the flowBDT model when applied to more structured datasets, such as point clouds that inherently contain spatial relationships and geometry.

\subsection{Conditional generation}\label{subsec4.3}

Most generative models encounter similar challenges when attempting to generate high-quality samples from non-Gaussian priors in large dimensions or from distributions that are significantly different from the target distribution. In this section, we will show how conditional generation tackled by the model in both low or high dimensions.

Due to the restricted resolution and acceptance of detectors, the distributions of detector-level quantities diverge from those at the particle level. One important technique used in collider physics is unfolding which corrects these detector effects. Machine learning-based unfolding methods \cite{cinnfold,sbfold,omnifold} offer significant advantages for unbinned and high-dimensional differential cross-section measurements.

The public jet dataset with several MC generator/tune pairs is used, we consider the leading jet width and mass, multiplicity, N-subjetiness ratio between leading and subleading jets $\tau_{21} = \frac{\tau_2}{\tau_1}$, the normalized groomed mass $\ln\rho = \frac{\ln m_{SD}^2}{p_T^2}$ and momentum fraction $z_g$ for Soft Drop grooming algorithm with specific angularity and cut \cite{softdrop0,softdrop1}. To assess the performance of the conditioned generation, we compare with the histograms generated from a Gaussian prior using both binned and unbinned metrics. As shown in Table.2, we found that the separation power and EMD from the non-Gaussian prior demonstrate better performance for the majority of the time. More importantly, starting from a reconstructed level prior provides substantial benefits, as it captures the correlation of the target observables much more effectively than a multidimensional Gaussian would. If we define a relative improvement on correlation difference to be ratio of the sum of absolute values of those differences with data between simulation and generated sample:
\begin{gather}
    \mathcal{I}_{rel} = \frac{\sum_{i=1}^n Corr_{data,i}-Corr_{sim,i}}{\sum_{i=1}^n Corr_{data,i}-Corr_{gen,i}}
\end{gather}
The relative improvement for conditional generation is around 3040\%, which is much higher than 414\% improvement calculated from unconditional one. But still both cases show a huge enhancement on general correlation as shown in Fig.~\ref{fig9}.

\begin{figure*}[htbp]
    \centering
    \includegraphics[width=\linewidth]{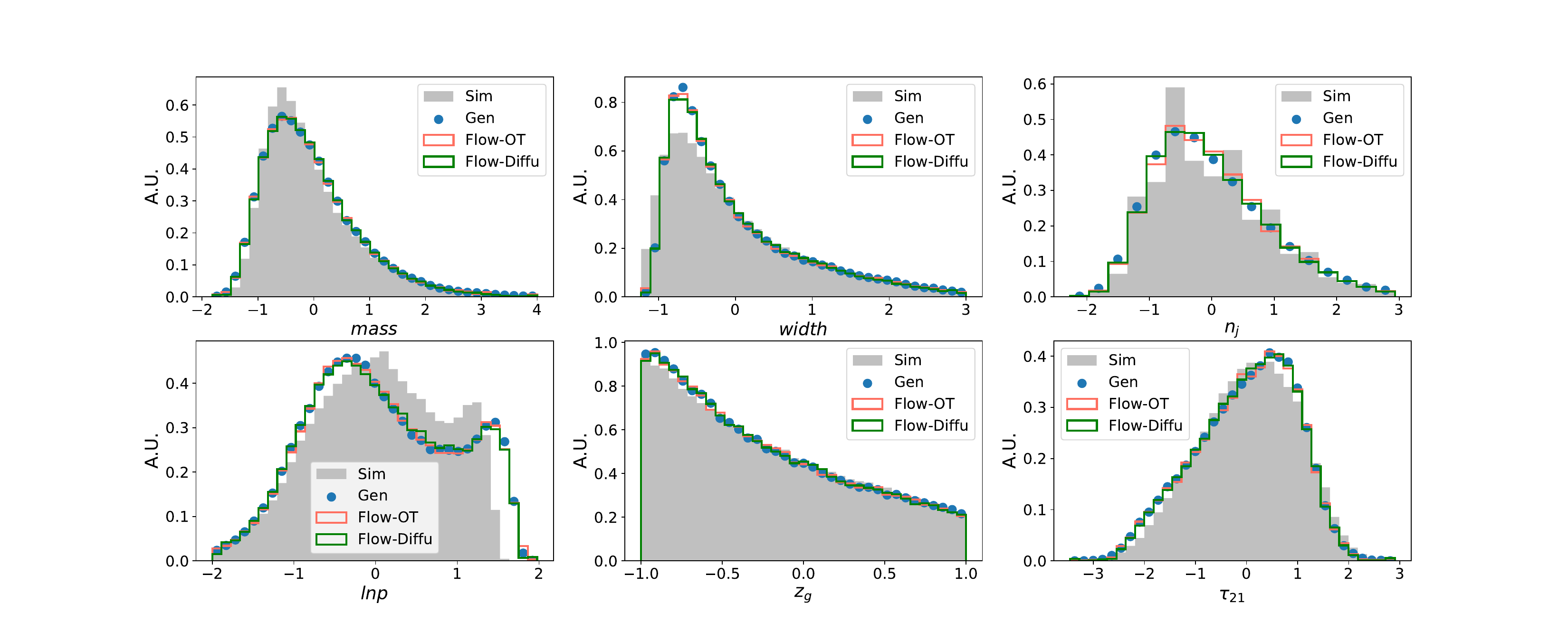}
    \caption{Comparison plots of the histograms for 6 observables from leading jets for unfolding. Flow-Diffu: from Gaussian prior, Flow-OT: from conditioned prior, Gen: target unfolded observables, Sim: from reconstructed events.}
    \label{fig8}
\end{figure*}

\begin{figure*}[htbp]
    \centering
    \includegraphics[width=\linewidth]{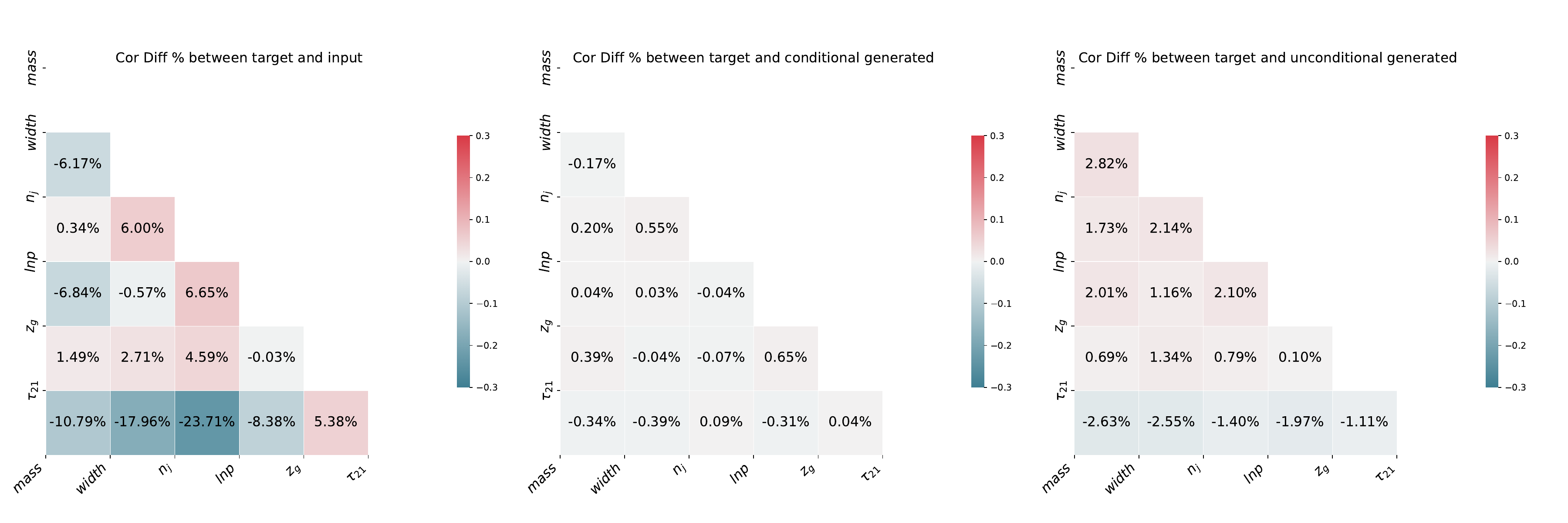}
    \caption{Comparison plots of the correlation difference heatmaps for 6 observables from leading jets for unfolding. Left: Between target data and original simulation level input, middle: Between target data and generation from conditioned prior, right: Between target data and generation from Gaussian prior. }
    \label{fig9}
    
\end{figure*}

\begin{table*}[htbp]
    \centering
    \begin{ruledtabular}
        \begin{tabular}{cccc}
        \textbf{Observables}(\textbf{Sep power}$\times$100 / \textbf{EMD}$\times$10) & \textbf{Simulation} & \textbf{Conditional} & \textbf{Unconditional} \\ \hline
        \textbf{Jet mass} &0.496/0.369 & \textbf{0.072}/\textbf{0.038} &0.091/0.084 \\
        \textbf{Jet width} &1.770/0.406 &0.044/\textbf{0.031} &\textbf{0.034}/0.098 \\
        \textbf{N constituents} &1.129/0.405 &0.10/0.28 &\textbf{0.07}/\textbf{0.26} \\
        \textbf{$\ln_{\rho}$} &3.684/0.916 &\textbf{0.053}/\textbf{0.046} &0.10/0.082 \\
        \textbf{$z_g$} &0.047/0.173 &\textbf{0.010}/\textbf{0.029} &0.013/0.030 \\
        \textbf{$\tau_{21}$} &0.558/0.593 &\textbf{0.095}/\textbf{0.054} &0.14/0.093 \\
            \end{tabular}
    \end{ruledtabular}
    
    \caption{Comparison table for evaluation metrics over binned and unbinned distributions of 6 unfolding observables between data and original simulation, generated sample with and without conditional inputs}
    \label{table1}
\end{table*}

In a shift of focus, the model is utilized to refine and simulate calorimeter showers. The conditional input is derived from parameterized shower approximations produced by fast simulations. In typical machine learning models, we often use discrete conditions such as energy and coordinates for training. However, some attributes, such as material type, are difficult to define numerically. Using an approximate shower as a condition offers a novel way to integrate these factors to guide the model with more explicit information inside the calorimeter.

\begin{figure}[htbp]
    \centering
    \includegraphics[width=0.4\textwidth]{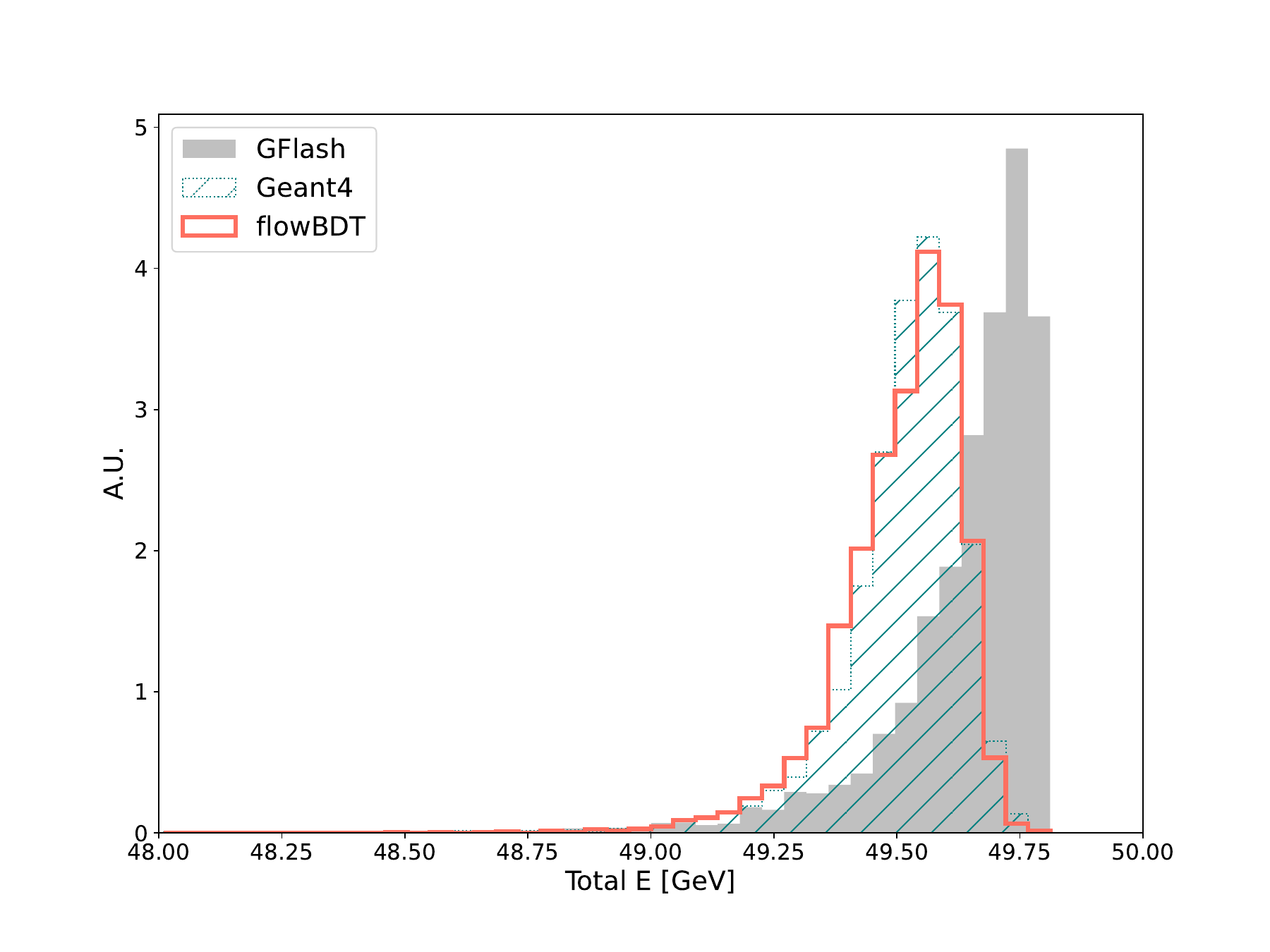}
    \caption{Total energy deposited for 50 GeV electron shower in the 10 * 10 calorimeter. GFlash: from classical fast simulation, flowBDT: generated sample, Geant4: from full simulation}
    \label{fig10}
\end{figure}

Again, we flatten the calorimeter cells before training. Although this increases the training time compared to tasks with fewer dimensions, the computational cost is still generally smaller than most models we have encountered. Figure~\ref{fig10} shows the histogram of the deposited energy sum for the electron shower. The generated histogram displays a peak width very similar to the original, but with sharper tails at the lower energy end.

\section{Conclusion and outlooks}\label{sec5}

In this work, we have introduced BUFF, a new framework with applications of flow matching and extensive usage of GBT models to accomplish various tasks on different levels of analysis in HEP. We have developed a novel model called flowBDT, which features several enhancements over the original, such as the use of higher-order solvers to reduce training time, distinct GBT training objectives to handle very high-dimensional tasks with a 5-8 times speedup in sampling, and the benefits of conditional generations. The negligible inference time of lightweight GBT models significantly shortens the generation time for high-level tasks compared to traditional flow matching. BUFF is indeed well-suited for rapid training across multiple CPU cores. 

We evaluated the generated high-level features for complicated top jet topology using both binned and unbinned metrics. The results, including the f-divergence metric and Earth Mover’s Distance, are impressive for most features.  We greatly increase the dimensionality for the simulation of low-level calorimeter cells and jet constituents. Our model still shows very decent result for those tasks. Then, we demonstrated some advantages of conditional generation on both high and low levels. For high level tasks such as unfolding, the relative improvement on correlation difference with target from conditional generations is over 30 times higher than before. The conditions can also be the cells from an approximate shower that takes geometry and material information into account. 

The model outperforms most tasks specific to tabular data than typical flow-based models. Looking ahead, we aim to extend the application of our model to include a variety of complex tasks such as anomaly detection, quark-gluon tagging, and others. Additionally, for low-level simulations, we will explore alternative strategies to further enhance the efficiency, aiming to significantly speed up both the training and inference time. 

\section{Acknowledgements}

We would like to thank Jie Xiao for his insightful inputs on the study of conditional generation, and Dominik Duda for discussion on realistic application scenarios under the context of the ATLAS experiment.

\begin{appendix}
\section{Sampling time metric}\label{secA1}

\begin{figure*} [htbp]
    \centering
        \centering
            \includegraphics[width=0.4\textwidth]{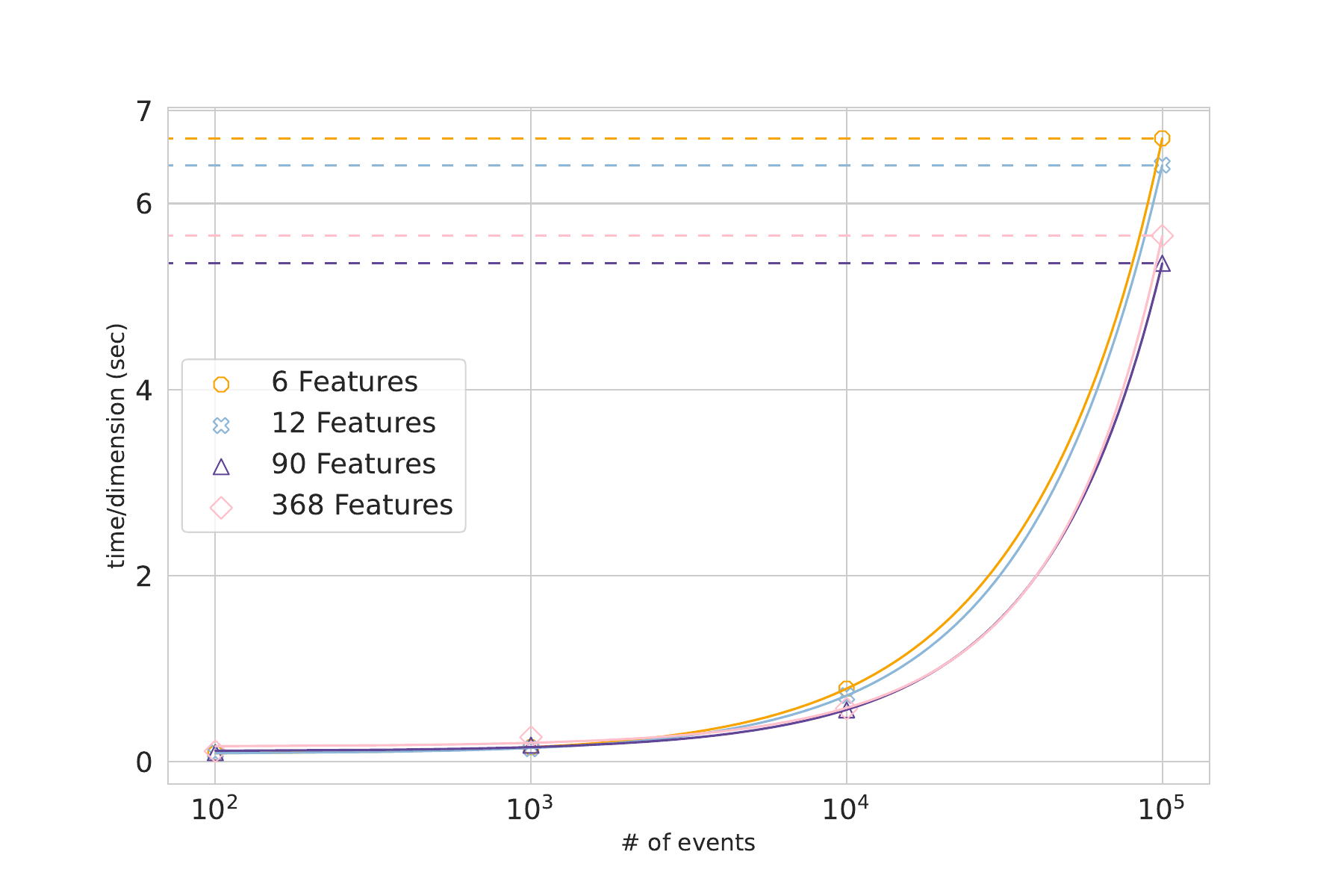}
            \quad\quad\quad
            \includegraphics[width=0.4\textwidth]{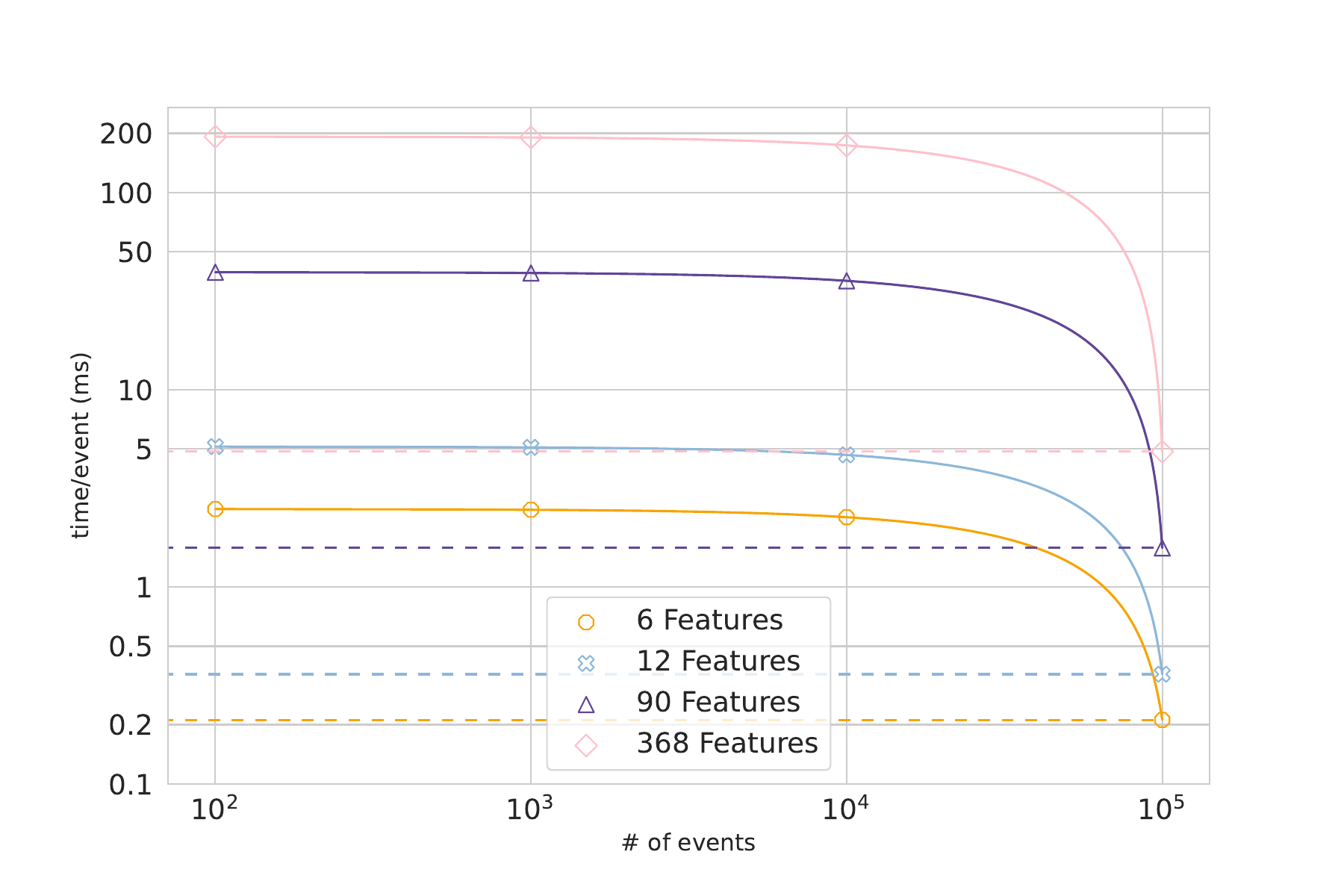}
        \caption{Sampling time per feature (left) and per event (right) for number of events generated at once with different dimensionalities.}
        \label{fig11}
\end{figure*}

We listed two plots in Fig.~\ref{fig11} to graphically demonstrate how should we choose the batch size of generated sample given different dimensonalities. Since multiple GBT models are used during sampling stage, each model needs a short period time to be "awake". This is a non-negligible factor if our sampling time is already below millisecond scale. Any small batch size of generated sample would often required a larger sampling time per feature or per event. For example, the total generation time only scales up to 2 times when increasing from 100 to 1000 events, and up to 4-5 times when scaling from 1000 to 10,000 events. This demonstrates the model's efficient scaling behavior, which is crucial for processing large datasets. Noticeably, for restricting the number of output on each tree would result in a faster inference, this is especially obvious for higher dimensions. As we see in the right plot in Fig.~\ref{fig11}, an event with 368 features only consumes 2.5 times more CPU cost than an event with 90 features.

\end{appendix}

\newpage

\bibliography{prd.bib}

\end{document}